\documentclass[prb,twocolumn,superscriptaddress,showpacs,floatfix,long	bibliography]{revtex4-2}
\usepackage{graphicx}
\usepackage{bm}
\usepackage{color,soul}
\usepackage{amsmath,amssymb,amsfonts,float,graphics,epsfig,epstopdf,color,verbatim,tabularx,bm,multirow,appendix,hyperref}

\begin{document}

\title{Neutron-diffraction and linear {Gr\"uneisen}  parameter studies of magnetism in NdFe$_2$Ga$_8$}
\author{Xingyu Wang}
\affiliation{Beijing National Laboratory for Condensed Matter Physics, Institute of Physics, Chinese Academy of Sciences, Beijing 100190, China}
\affiliation{School of Physical Sciences, University of Chinese Academy of Sciences, Beijing 100190, China}
\author{Cuixiang Wang}
\affiliation{Beijing National Laboratory for Condensed Matter Physics, Institute of Physics, Chinese Academy of Sciences, Beijing 100190, China}
\affiliation{Center of Materials Science and Optoelectronics Engineering, University of Chinese Academy of Sciences, Beijing 100049, China}
\affiliation{School of Physical Sciences, University of Chinese Academy of Sciences, Beijing 100190, China}
\author{Bo Liu}
\affiliation{Beijing National Laboratory for Condensed Matter Physics, Institute of Physics, Chinese Academy of Sciences, Beijing 100190, China}
\affiliation{School of Physical Sciences, University of Chinese Academy of Sciences, Beijing 100190, China}
\author{Ke Jia}
\affiliation{Beijing National Laboratory for Condensed Matter Physics, Institute of Physics, Chinese Academy of Sciences, Beijing 100190, China}
\affiliation{School of Physical Sciences, University of Chinese Academy of Sciences, Beijing 100190, China}
\author{Xiaoyan Ma}
\affiliation{Beijing National Laboratory for Condensed Matter Physics, Institute of Physics, Chinese Academy of Sciences, Beijing 100190, China}
\affiliation{School of Physical Sciences, University of Chinese Academy of Sciences, Beijing 100190, China}
\author{Gang Li}
\affiliation{Beijing National Laboratory for Condensed Matter Physics, Institute of Physics, Chinese Academy of Sciences, Beijing 100190, China}
\affiliation{Songshan Lake Materials Laboratory , Dongguan, Guangdong 523808, China}
\author{Xiaoping Wang}
\affiliation{Neutron Scattering Division, Oak Ridge National Laboratory, Oak Ridge, Tennessee 37831, USA}
\author{Chin-Wei Wang}
\affiliation{Neutron Group, National Synchrotron Radiation Research Center, Hsinchu 30076, Taiwan}
\author{Youguo Shi}
\email{ygshi@iphy.ac.cn}
\affiliation{Beijing National Laboratory for Condensed Matter Physics, Institute of Physics, Chinese Academy of Sciences, Beijing 100190, China}
\affiliation{Center of Materials Science and Optoelectronics Engineering, University of Chinese Academy of Sciences, Beijing 100049, China}
\affiliation{Songshan Lake Materials Laboratory , Dongguan, Guangdong 523808, China}
\author{Yi-feng Yang}
\email{yifeng@iphy.ac.cn}
\affiliation{Beijing National Laboratory for Condensed Matter Physics, Institute of Physics, Chinese Academy of Sciences, Beijing 100190, China}
\affiliation{School of Physical Sciences, University of Chinese Academy of Sciences, Beijing 100190, China}
\affiliation{Songshan Lake Materials Laboratory , Dongguan, Guangdong 523808, China}
\author{Shiliang Li}
\email{slli@iphy.ac.cn}
\affiliation{Beijing National Laboratory for Condensed Matter Physics, Institute of Physics, Chinese Academy of Sciences, Beijing 100190, China}
\affiliation{School of Physical Sciences, University of Chinese Academy of Sciences, Beijing 100190, China}
\affiliation{Songshan Lake Materials Laboratory , Dongguan, Guangdong 523808, China}

\begin{abstract}
We study the magnetism in NdFe$_2$Ga$_8$ by the neutron-diffraction and temperature-modulated linear {Gr\"uneisen}  parameter measurements. Previous thermodynamical measurements have demonstrated that there are two magnetic transitions at 10 and 14.5 K, respectively. Neutron-diffraction measurements confirm that the lower one is an antiferromagnetic (AFM) transition with a commensurate magnetic structure. Both the commensurate and the incommensurate (IC) magnetic peaks are found below the higher transition but their intensities only gradually increase with decreasing temperature. Below 10 K, the commensurate peak intensity increases quickly with decreasing temperature, signaling the AFM transition, while the IC peak intensity disappears below 5 K. The linear {Gr\"uneisen} parameter along the $c$ axis, $\Gamma_c$, shows a hysteresis behavior that is different from the hysteresis behavior for the magnetization $M$. We give a discussion of the origin of the magnetism in NdFe$_2$Ga$_8$.
\end{abstract}



\maketitle

\section{introduction}

The study of heavy-fermion materials has long been one of the important fields in condensed matter physics. Recently, the so-called 1-2-8 system, i.e., $Ln$$M$$_2$$T$$_8$ ($Ln$ = La, Nd, Ce, etc., $M$ = Fe, Co, Ru, etc. and $T$ = Al, Ga, In), has attracted increasing interests \cite{OgunbunmiMO20}. These materials have the CaCo$_8$Al$_8$-type structure with the space group of $Pbam$ with $L_n$ ions forming one-dimensional (1D) chains along the $c$ axis \cite{SichevichOM85}. The magnetic properties are mainly determined by the rare-earth elements as in many other heavy-fermion materials. For example, an antiferromagnetic (AFM) order is found in Pr-based and Eu-based compounds \cite{TougaitO05,NairHS16,NairHS17,OgunbunmiMO20b,OgunbunmiMO17,OgunbunmiMO18,WatkinsCP15,FritschV04,SichevychO09,CaltaNP16}. A ferromagnetic order is found in CePd$_2$Al$_8$ with $T_c$ = 9.5 K, but it has a monoclinic crystal structure \cite{TursinaA18}. For the Ce-based compounds with the orthorhombic structure, no magnetic ordering is found \cite{KolendaM01,SchluterM01,GhoshS12,ChengK19,WangL17,BhattacharyyaA20}. The CeCo$_2$Ga$_8$ is shown to be a rare example of a quasi-1D Kondo chain \cite{ChengK19},  but its low-temperature properties show 2D quantum critical behaviors including the temperature dependence of the magnetic susceptibility ($\chi \sim T^{-0.2}$), the specific heat ($C/T \sim -lnT$) and the resistivity ($\rho \sim T^n$ with $n \sim$ 1), which may be associated with a quantum critical point \cite{WangL17,BhattacharyyaA20}.

The Nd-based 1-2-8 compounds have been relatively less studied. The NdCo$_2$Al$_8$ exhibits an AFM order at 8.7 K \cite{HeW09,WatkinsCP15}. Interestingly, NdFe$_2$Ga$_8$ has two magnetic transitions at about 15 and 10 K \cite{WangC21}, which is unique in the 1-2-8 compounds.  Both transitions are hardly affected by the magnetic field within the $ab$ plane, but they can be quickly suppressed by the field along the $c$ axis. At 7 T where magnetic transitions are completely suppressed, the system shows quantum criticality of the conventional 3D spin-density-wave type, such as $\rho \sim T^{1.5}$  and $C/T \sim T^{-0.5}$. While both transitions have been assumed to be AFM in nature, the magnetic structures have not been identified. Moreover, a hysteresis behavior is found in the magnetic-susceptibility and resistivity measurements, and its origin is still unknown. 

In this work, we studied the magnetic structure of NdFe$_2$Ga$_8$ by the neutron-diffraction technique. We found that the transition at $T_N \sim $ 10 K is from a long-range commensurate AFM order. Below the higher transition temperature $T_O \sim$ 14.5 K, both incommensurate (IC) and commensurate magnetic peaks gradually appear. The intensities of the IC peaks become maximum at $T_N$ and decrease to zero at about 5 K. The field dependence of the linear {Gr\"uneisen} parameter along the $c$ axis, $\Gamma_c$, shows several maximums and a hysteresis behavior that is different from that of the magnetization $M$. Our results suggest that the magnetic orders are complicated. 

\section{experiments}

Single crystals of NdFe$_2$Ga$_8$ were grown by the self-flux method as reported previously \cite{WangC21}. Neutron single-crystal diffraction (NSCD) measurement was carried out on a 113-mg single-crystal sample on the TOPAZ diffractometer at the Spallation Neutron Source, Oak Ridge National Laboratory. Five grams of single crystals were ground to powders and measured by the neutron powder diffraction (NPD) on the Echidna diffractometer at the Australian Nuclear  Science and Technology Organization. The refinements for the NSCD and NPD data were done by using the Jana and Fullprof programs, respectively \cite{PetricekV14,RodriguezJ93}. The linear {Gr\"uneisen} parameter was measured by a home-made temperature-modulated dilatometer \cite{GuY20}  putting in a physical property measurement system (Quantum Design) and a 18-T magnet with a top-loading $^{3}$He fridge. The dimension of the sample is about 0.2 mm in diameter and 4 mm in length. The sample was attached on the top of the piezobender and CuBe frame by glue, and self-heated by two silver-paint contacts. Several heating currents have been tested to make sure that the measurement is in the optimal condition \cite{GuY20}.

\section{results and discussions}

\begin{figure}[tbp]
\includegraphics[width=\columnwidth]{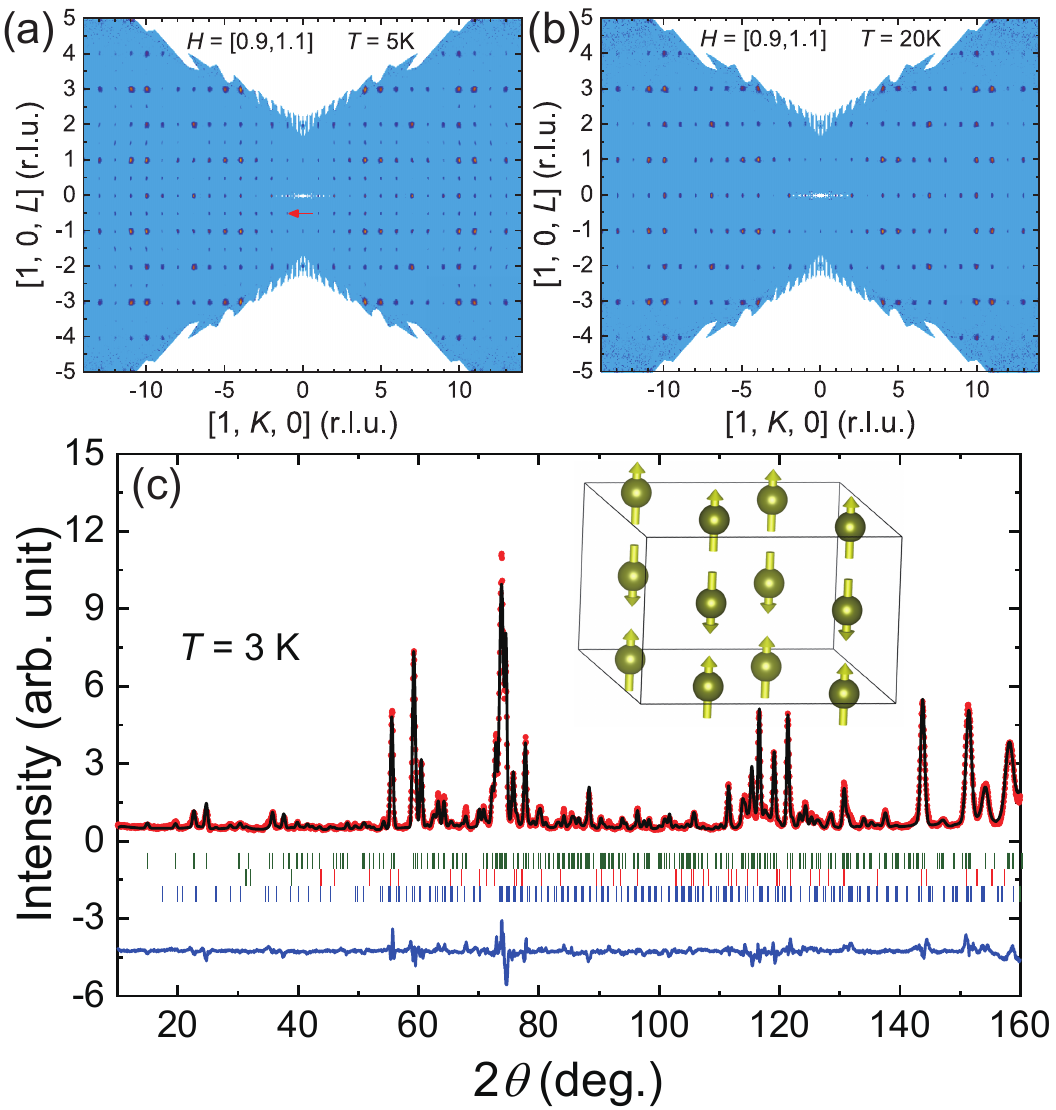}
 \caption{(a) and (b) NSCD patterns in the (1,$K$,$L$) plane at 5 and 20 K, respectively. The brackets for $H$ give the integrated range along the $H$ direction. The arrow in (a) indicates a new peak appeared at half integer $L$. (c) NPD intensity at 3 K. The calculated intensities are shown by the black line. Short vertical
green and blue lines represent calculated nuclear and magnetic Bragg peak positions, respectively. Vertical red lines represent peaks from FeGa$_3$ impurities. The blue line shows the difference between measured and calculated intensities. The weighted profile R-factor ($R_{wp}$) is 10.2. The inset shows the magnetic structure with the moments sitting on Nd sites.}
\label{ND}
\end{figure}

Figure \ref{ND}(a) and \ref{ND}(b) show the NSCD patterns of NdFe$_2$Ga$_8$ in the [1,$K$,$L$] plane at 5 and 20 K, respectively. Compared with 20-K data, new peaks appear at 5 K at half integer $L$, suggesting the appearance of long-range AFM order. Nuclear structure was refined at 20 K with 23033 peaks and the result is the same as previously reported \cite{WangC21}. At 5 K, no change of nuclear structure is found and the magnetic structure is refined by the 5231 magnetic peaks with a total wR of 26.36. Figure \ref{ND}(c) shows the NPD data at 3 K and the refinement for both nuclear and magnetic structures gives the magnetic R factor as 60.5. About 3.2\% of FeGa$_3$ impurity is found in the NPD pattern. We note that this impurity does not affect the NSCD refinement as it has the form of polycrystalline and its diffraction intensity in the single-crystal sample is thus very weak. In both refinements, the magnetic form factor has included both $<j_0>$ and $<j_2>$ terms. Both the single-crystal and the polycrystalline refinements give the same magnetic structure as shown in the inset of Fig. \ref{ND}(c).  The magnetic moments point to the $c$ axis, and are aligned ferromagnetically within the ab plane but antiferromagnetically along the c axis. The value of the ordered moment from the NSCD and NPD is 3.379 $\pm$ 0.016 $\mu_B$ at 5 K and 3.31 $\pm$ 0.15 $\mu_B$ at 3 K, respectively. 

We note that the above refinements have assumed that the moments are on the Nd sites. If the moments are added on Fe sites, the value on Fe1 site is negligible, while that on the Fe2 site is about 0.593(29) and 0.544(27) $\mu_B$ for Fe$^{2+}$ and Fe$^{3+}$ magnetic form factors used in refinements, respectively, which are much smaller than the moments on Nd sites. However, we find no improvement to the fit with the Fe moments included and the Nd moments are almost the same as before. Therefore we conclude that Nd$^{3+}$ is the only magnetic ion in the magnetic ordering of NdFe$_2$Ga$_8$. A similar conclusion has also been found in PrFe$_2$Al$_8$ \cite{NairHS17}.

\begin{figure}[tbp]
\includegraphics[width=\columnwidth]{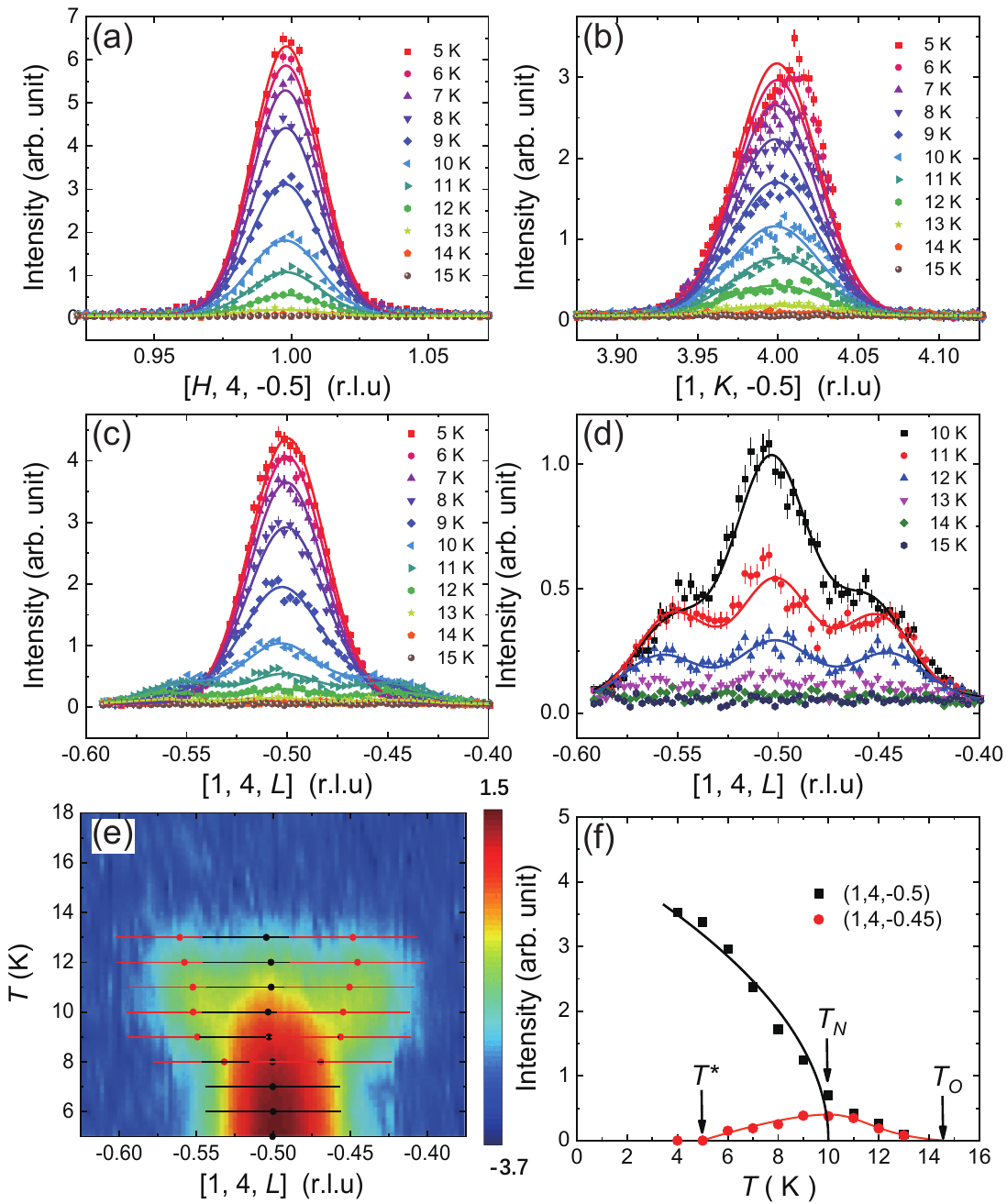}
 \caption{(a) and (b) The cuts at (1,4,-0.5) peak from the NSCD data at different temperatures along the $H$ and $K$ directions, respectively. The solid lines are fitted by the Gaussian function. (c) The cuts at (1,4,-0.5) peak along the $L$ direction at different temperatures. The solid lines are fitted by the three-Gaussian function. Panel (d) shows the data in (c) for temperatures from 10 to 15 K. (e) The colormap for the intensity at (1,4,-0.5) in logarithmic scale. The dots mark the centers of the commensurate and IC peaks. The horizontal bars give the FWHMs of the peaks. (f) The temperature dependence of the peak intensity at (1,4,-0.5) and (1,4,-0.45). The black solid line is fitted by the function as described in the text. The red line is a guide to the eye. }
\label{IC}
\end{figure}

Figures \ref{IC}(a) and \ref{IC}(b) show the magnetic Bragg peaks at (1,4,-0.5) cutting along the $H$ and $K$ directions, respectively. With the peak intensity decreasing with increasing temperature, the peak can always be well fitted by the Gaussian function with no change of the width. For the cuts along the $L$ direction [Fig. \ref{IC}(c)], while the peak can still be fitted by the Gaussian function at low temperatures, two new peaks emerge at the IC positions ($L \sim -0.5 \pm \delta$) with $\delta \approx$ 0.05 at high temperatures.  A three-Gaussian function with the sum of three individual Gaussian functions is thus introduced to fit the data, as shown in Fig. \ref{IC}(d). Since we are unable to refine the magnetic structure for the IC peaks, it is not known whether Fe may play a role in this phase or not.

Figure \ref{IC}(e) shows the colormap for the intensity as a function of $L$ and temperature. The incommensurability $\delta$ varies little above 10 K and seems to decrease with decreasing temperature below 10 K. The full width at half maximums (FWHMs) for the commensurate and IC peaks are both temperature independent and comparable to the instrumental resolution, indicating that both of the peaks are associated with long-range orders.

Figure \ref{IC}(f) shows the temperature dependence of the peak intensities at the commensurate and IC positions. Below 10 K, the commensurate one can be fitted by the function $I_0(1-T/T_N)^{2\beta}$ with $T_N$ fixed 10 K. The fitted $\beta$ is 0.31 $\pm$ 0.03. The IC peak intensity shows totally different behaviors. Above $T_N$, the temperature dependence of both commensurate and IC peak intensities is the same. Within our resolution, we can only identify the appearance of magnetic peaks at 13 K. With decreasing temperature, the IC peak intensity reaches a maximum at $T_N$ and decreases with further cooling temperature, whereas the commensurate peak intensity quickly increases as a regular AFM transition. The IC peaks completely disappear below $T^* \approx$ 5 K. This suggests that the IC phase may compete with the low-temperature commensurate magnetic phase.

\begin{figure}[tbp]
\includegraphics[width=\columnwidth]{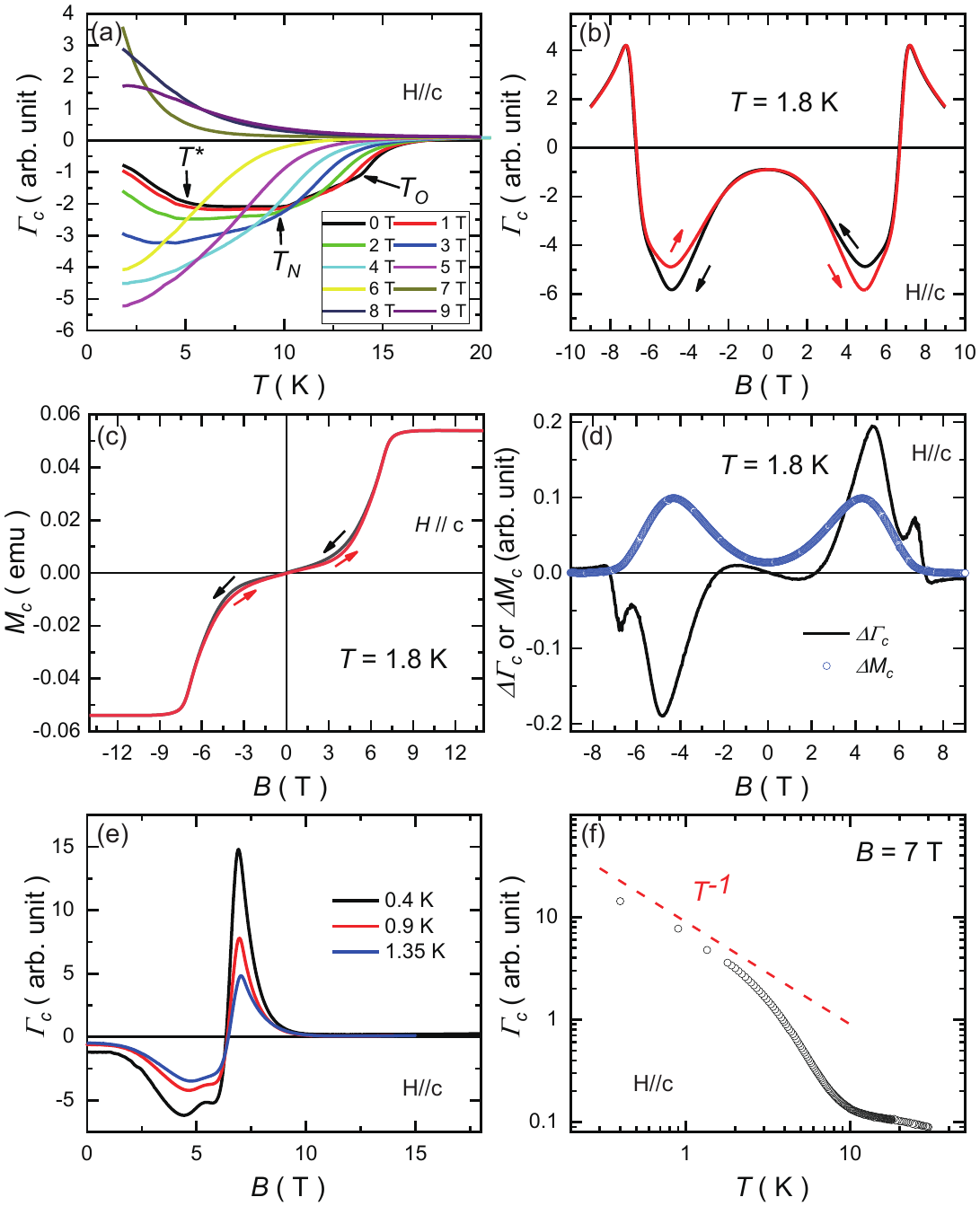}
 \caption{(a) Temperature dependence of the linear {Gr\"uneisen} parameter $\Gamma_c$ at various fields. The measurements were made in the FC process. The arrows indicate $T_O$ and $T_N$ at zero field. (b) Field dependence of $\Gamma_c$ at 1.8 K. The arrows indicate the field-decreasing and field-increasing processes. (c) Field dependence of $M_c$ at 1.8 K. The arrows indicate the field-decreasing and field-increasing processes. (d) Field dependence of $\Delta\Gamma_c$ and $\Delta M_c$ at 1.8 K.  (e) Field dependence of $\Gamma_c$ below 1.8 K for a different sample. The measurements were made with the field-decreasing process. (f) $\Gamma_c$ as a function of temperature at 7 T in the log-log scale. The values below 1.8 K are taken from the data in (e) and normalized to those above 1.8 K. The dashed line represents the $T^{-1}$ dependence.}
\label{Gamma}
\end{figure}

To further investigate the nature of the magnetic transitions in NdFe$_2$Ga$_8$, we studied the linear {Gr\"uneisen} parameter $\Gamma_c$, where the linear thermal expansion is measured along the $c$ axis \cite{GuY20}. Figure \ref{Gamma}(a) shows the temperature dependence of $\Gamma_c$ in the field-cooling (FC) process, which is small and positive at high temperatures for all fields. At zero field, $\Gamma_c$ changes to negative with decreasing temperature and its absolute value quickly increases until $T = T_O$. With a kink at $T_O$, $|\Gamma_c|$ continues to increase with decreasing temperature and shows another kink at $T_N$. Below $T_N$, $|\Gamma_c|$ initially changes little with decreasing temperature and then quickly decreases with decreasing temperature below $T^*$. With field increasing, $T_O$ decreases and seems to merge together with $T_N$ above 3 T, which is consistent with previous reports \cite{WangC21}. The round feature associated with $T^*$ decreases with the increasing field and cannot be seen above 3 T. Further increasing the field makes it hard to identify $T_N$ and $T_O$, while $|\Gamma_c|$ at 1.8 K becomes maximum at 5 T and starts decreasing with increasing field. Interestingly, $\Gamma_c$ at low temperature changes continuously from negative to positive value when the field increases from 6 to 7 T.

Figure \ref{Gamma}(b) shows the field dependence of $\Gamma_c$ at 1.8 K, where the positive and negative maximums appear at about 7 and 5 T, respectively. Figure \ref{Gamma}(c) shows the field dependence of $M_c$ with the field parallel to $c$ at 1.8 K. $M_c$ becomes fully saturated above about 9 T. Hysteresis behaviors are found between increasing and decreasing field processes for both measurements. Figure \ref{Gamma}(d) further shows the field dependence of $\Delta\Gamma_c$ = $\Gamma_c^{dec} - \Gamma_c^{inc}$, which is the difference of $\Gamma_c$ between the decreasing and the increasing field processes. As a comparison, the field dependence of $\Delta M_c$ = $M_c^{dec}-M_c^{inc}$ is also shown. Surprisingly, there are two maximums in $\Delta\Gamma_c$ at about 4.8 and 6.8 T, but only one maximum exists in $\Delta M_c$ at about 4.3 T. Moreover, the value of $\Delta\Gamma_c$ is initially small and changes sign at about 2.2 T. Its value increases dramatically above 3 T, while $\Delta M_c$ smoothly increases with increasing field until it reaches its maximum at 4.3 T. These results suggest that the hysteresis behaviors in $\Gamma_c$ and $M_c$ have different origins.

Figure \ref{Gamma}(e) shows $\Gamma_c$ in the field-decreasing process down to 0.4 K. Interestingly, two-peak features appear around 5 T at 0.4 K, which indicates complicated behaviors within the ordering state. We note that the transitions at $T_O$ and $T_N$ behave very differently in the field in that the former is easily suppressed by the field whereas the latter is little affected by the field below 3 T where the two transitions merge together \cite{WangC21}. The features around 5 T may thus be related to the couplings between the orders associated with $T_O$ and $T_N$. Nevertheless, here we are more interested in the properties around 7 T, which has been shown to exhibit quantum criticality \cite{WangC21}. Figure \ref{Gamma}(f) shows the temperature dependence of $\Gamma_c$ at 7 T, which exhibits a divergent behavior with $\Gamma_c \propto T^{-1}$ below 2 K. 

With the results above, we can get a more comprehensive understanding on the magnetic orderings in NdFe$_2$Ga$_8$. The magnetic transition at $T_N \sim$ 10 K is clearly associated with the commensurate AFM structure. Between $T_N$ and $T_O$, one commensurate and two IC peaks are observed. The IC magnetic structure may be explained as a SDW order \cite{WangC21}. This IC magnetism seems to compete with the commensurate order as the intensities of the IC peaks gradually decrease with decreasing temperature below $T_N$ and finally become zero below $T^*$ [Fig. 2(f)]. This competition seems to also present under field as $T^*$ is suppressed under field [Fig. 3(a)], which suggests that the IC peaks may be evoked by the magnetic field. 

There are some interesting features that need to be further discussed. First, the emergence of the IC peaks below $T_O$ does not look like the temperature dependence of a typical order parameter while the specific-heat jump at $T_O$ clearly indicates that the transition is a second-order transition. By forcing a fit of $I_0(1-T/T_O)^{2\beta}$ for the IC peak intensity above $T_N$, we get a value of $\beta$ much larger than 0.5, which is hard to understand. Second, $\Delta\Gamma_c$ shows two peaks whereas $\Delta M_c$ shows just one, as shown in Fig. \ref{Gamma}(d). The high-field peak of $\Delta\Gamma_c$ is located at the field around which $\Gamma_c$ changes quickly from negative value to positive value, which is typically associated with a dramatic change of the electronic state. This peak cannot be found in $\Delta M_c$, suggesting that it is not related to the dipole moments. As discussed above, the competition between the commensurate and IC magnetism may present under magnetic field, but it can not completely explain the hysteresis in $M - H$ and $\Gamma_c - H$ since they have different behaviors. 

The above results suggest that the magnetic phase below $T_O$ may not be a conventional SDW as expected previously \cite{WangC21}. One possibility is that the system is close to a Lifshitz point \cite{HornreichRM80} where the AFM and modulated phases could coexist but whether this scenario can explain the above results is not known. We propose another possibility that the IC phase may be a MO \cite{SantiniP09,SuzukiMT18}, which gives a natural explanation on why the temperature dependence of the IC peaks does not look like an order parameter. In this case, the hysteresis of $\Gamma_c$ is determined by the field dependence of MO or its competition with the long-range AFM order. We note that a MO could in principle result in large magnetic anisotropy and IC magnetic peaks. If a MO indeed exists in NdFe$_2$Ga$_8$, the itinerancy of electrons should play a major role as shown by the behaviors of the resistivity and  linear {Gr\"uneisen} parameter. Of course, there is still a lack of direct evidence for the existence of a MO and further studies on this system are needed.

\section{conclusions}

In conclusion, we have found both commensurate and IC magnetic peaks in the magnetic ordered states of NdFe$_2$Ga$_8$. While the low-temperature commensurate peaks are associated with the AFM transition at 10 K, the IC peaks cannot be simply understood within the picture of a conventional SDW. The measurements on the linear {Gr\"uneisen} parameter suggests that its behaviors may not be related to the dipole moments. While there are possibilities that theories based on the dipole moments may explain this, we suggest that a MO with strong itinerant characteristics could provide a simple answer.

\begin{acknowledgments}

This work is supported by the National Key R\&D Program of China (Grants No. 2020YFA0406003, No. 2017YFA0302900, No. 2017YFA0303100, No. 2016YFA0300502, and No. 2016YFA0300604), the National Natural Science Foundation of China (Grants No. 11874401, No. 11674406, No. 11961160699, No. 11774399, and No. U2032204), the Strategic Priority Research Program of the Chinese Academy of Sciences (Grant No. XDB33010000), the Beijing Natural Science Foundation (Z180008), and the K. C. Wong Education Foundation (GJTD-2020-01, GJTD-2018-01). The single-crystal neutron diffraction experiment on TOPAZ used resources at the Spallation Neutron Source, a DOE Office of Science User Facility operated by the Oak Ridge National Laboratory, USA.

\end{acknowledgments}

\end{document}